\documentclass[aps,prl,twocolumn,a4paper]{revtex4}
\pdfoutput=1

\newcommand{\kB}{k_{\textrm{B}}}
\newcommand{\kD}{k_{\textrm{D}}}
\newcommand{\kF}{k_{\textrm{F}}}
\newcommand{\UD}{U_{\textrm{D}}}
\newcommand{\tD}{t_{\textrm{d}}}
\newcommand{\kf}{k_{\textrm{cf}}}
\newcommand{\ts}{t_{\textrm{s}}}

\usepackage{geometry}
\usepackage{color}
\usepackage[latin1]{inputenc}
\usepackage{graphicx}
\usepackage{amsmath,amssymb}

\usepackage{times}

\usepackage{textcomp}
\usepackage{amsfonts}
\usepackage{graphicx}
\usepackage{bm}
\usepackage{color}
\usepackage[latin1]{inputenc}
\usepackage{braket}
\usepackage{units}
\usepackage[svgnames]{xcolor}
\usepackage[]{hyperref}

\def\be{\begin{equation}}
\def\ee{\end{equation}}

\begin{document}

\title{Synthetic dissipation and cascade fluxes in a turbulent quantum gas}

\author
{Nir Navon$^{1,2}$\footnote{To whom correspondence should be addressed: nir.navon@yale.edu, fujimoto@cat.phys.s.u-tokyo.ac.jp.},
	Christoph Eigen$^{1}$, 
	Jinyi Zhang$^{1}$, 
	Raphael Lopes$^{1}$\footnote{Present address: Laboratoire Kastler Brossel, Coll{\`e}ge de France, CNRS, ENS-PSL University,
		UPMC-Sorbonne Universit{\'e}, 11 Place Marcelin Berthelot, F-75005 Paris, France},
	Alexander L. Gaunt$^{1,3}$, 
	Kazuya Fujimoto$^{4\ast}$, 
	Makoto Tsubota$^{5,6}$, 
	Robert P. Smith$^{1,7}$ and Zoran Hadzibabic$^{1}$}

\affiliation{\vspace{1mm} $^1$ Cavendish Laboratory, University of Cambridge, J.J. Thomson Avenue, Cambridge CB3 0HE, United Kingdom\\
	$^2$ Department of Physics, Yale University, New Haven, Connecticut 06520, USA\\
	$^3$ Microsoft Research, 21 Station Road, Cambridge CB1 2FB, United Kingdom\\
	$^4$ Department of Physics, University of Tokyo, 7-3-1 Hongo, Bunkyo-ku, Tokyo 113-0033, Japan \\
	$^5$ Department of Physics, Osaka City University, 3-3-138 Sugimoto, Sumiyoshi-Ku, Osaka 558-8585, Japan \\
	$^6$ The OCU Advanced Research Institute for Natural Science and Technology, Osaka City University, Japan \\
	$^7$ Clarendon Laboratory, University of Oxford, Oxford OX1 3PU, United Kingdom
}

\begin{abstract}
Scale-invariant fluxes are the defining property of turbulent cascades, but their direct measurement is a notorious problem. Here we perform such a measurement for a direct energy cascade in a turbulent quantum gas. Using a time-periodic force, we inject energy at a large lengthscale and generate a cascade in a uniformly-trapped Bose gas. The adjustable trap depth provides a high-momentum cutoff $\kD$, which realises a synthetic dissipation scale. This gives us direct access to the particle flux across a momentum shell of radius $\kD$, and the tunability of $\kD$ allows for a clear demonstration of the zeroth law of turbulence: we observe that for fixed forcing the particle flux vanishes as $\kD^{-2}$ in the dissipationless limit $\kD\rightarrow \infty$, while the energy flux is independent of $\kD$. 
Moreover, our time-resolved measurements give unique access to the pre-steady-state dynamics, when the cascade front propagates in momentum space.
\end{abstract}

\maketitle

\begin{figure}[b]
	\centerline{\includegraphics[width=\columnwidth]{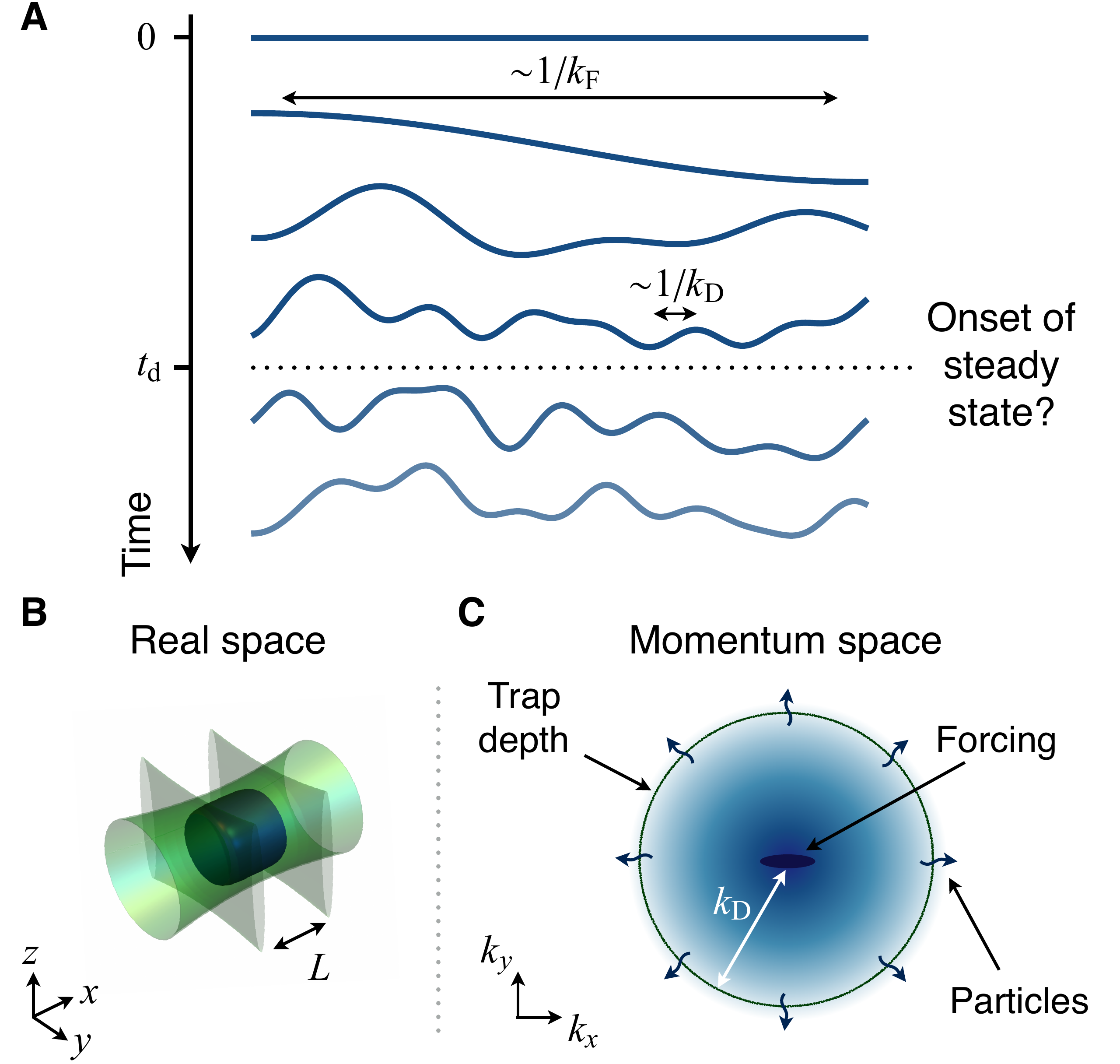}}
	\caption{{\bf Turbulent cascade in a box-trapped quantum gas.} ({\bf A}) Cartoon of real-space dynamics of a turbulent wave. Energy is injected by forcing the matter-wave field at a large lengthscale, $1/\kF$, and propagates to smaller scales due to nonlinear interactions. A steady state can be established once the excitations first reach the small dissipation lengthscale, $1/\kD$, at a time $\tD$. ({\bf B}) Sketch of the experimental setting. The atoms (blue) are trapped in a finite-depth potential formed by laser barriers (green) in the shape of a cylindrical box. The shaking force is applied along $\hat{\bf x}$.  ({\bf C}) In momentum space, the dissipation scale $\kD$ is set by the trap depth; when excitations propagate to $\kD$, dissipation occurs in the form of particle loss.
	} \label{Fig1}
\end{figure}

The discovery in 1941 by Kolmogorov and Obukhov of a universal law describing the transfer of energy from large to small lengthscales in turbulent flows was a conceptual breakthrough~\cite{kolmogorov1941local,obukhov1941distribution}. Despite their complex spatiotemporal dynamics, turbulent flows often obey a simple generic picture: the energy injected into the system at a large lengthscale flows locally in Fourier space, through lengthscales in the so-called inertial range where no dissipation occurs, until it is dissipated at some small lengthscale. In Fig.~\ref{Fig1}A, we depict such turbulent-cascade dynamics for a compressible field in real space.  
Here, a field initially at rest is at times $t>0$ continuously forced at a large lengthscale $1/\kF$, and the excitations propagate to smaller lengthscales due to nonlinear interactions. 
Once the excitations first reach the dissipation scale $1/\kD$, at time $\tD$, the field fluctuates on all lengthscales from $1/\kF$ to $1/\kD$. If a steady state is established within the momentum range $\kF$ to $\kD$, from thereon energy is dissipated at $\kD$ at the same rate at which it is injected at $\kF$. 
In such a steady state, the momentum-space distributions of quantities such as the energy or wave amplitude, are generically scale-free power laws.

Many quantitative theoretical predictions about turbulence are based on taking the mathematical limits $\kF\rightarrow 0$ and $\kD\rightarrow\infty$ \cite{zakharov1992kolmogorov}. 
Such formal treatments lead to predictions that are elegant, but often also counter-intuitive. A key prediction of this kind is that for $\kD\rightarrow\infty$ the steady-state cascade corresponds to a scale-invariant ($k$-independent) energy flux through momentum space, but no particle flux~\cite{dyachenko1992optical}.

Experimentally, the steady-state power-law spectra of various quantities have been extensively studied~\cite{frisch1995turbulence,alexandrova2009universality,salort2010turbulent,chepurnov2015turbulence,navon2016emergence}, while the equally fundamental cascade fluxes are harder to measure~\cite{mccomb2014homogeneous,uberoi1963energy,miyake1970airborne,deike2014energy}. 
Recently, ultracold atomic gases have emerged as a novel platform for studies of turbulence~\cite{henn2009emergence,neely2013characteristics,kwon2014relaxation,Tsatsos:2016,navon2016emergence,kwon2016observation,seo2017observation,mossman2017turbulence,johnstone:2018,gauthier2018negative}, which offers new experimental possibilities. 
Here, we use an atomic gas to directly measure cascade fluxes in a turbulent system. Moreover, our dissipation scale is tuneable, which allows us to explore how the fluxes depend on $\kD$, and to reconcile the experimental observations with the formal predictions for $\kD\rightarrow \infty$. 
Our system also allows a time-resolved study of the initial stage of turbulence, when steady state is not yet established, which reveals how the cascade front propagates in momentum space.

Our experiment starts with a quasi-pure weakly interacting Bose-Einstein condensate of $N\approx 1.2\times 10^5$ atoms of $^{87}$Rb in the uniform potential of a cylindrical optical box trap of radius $R\approx 16~\mu$m and length $L\approx 27~\mu$m (see Fig.~\ref{Fig1}B and~\cite{gaunt2013bose}). The chemical potential of the gas is $\mu\approx \kB \times 2$~nK, corresponding to a healing length $\xi\approx 1.2 $ $\mu$m $\ll$ $R$, $L$.
As in \cite{navon2016emergence}, we initiate a turbulent cascade by injecting energy at the system-size lengthscale, using a spatially uniform force ${\bf F}_\textrm{s}({\bf r},t)= F_0\sin(\omega_{\rm s} t)\hat{\bf x}$; here $\hat{\bf x}$ is a unit vector along the box symmetry axis, $F_0 L \approx\kB \times 2.5$~nK, and $\omega_{\rm s} \approx 2\pi \times 9$~Hz is tuned to resonantly excite the sound wave of wavelength $2L$ (so $\kF=\pi/L$)~\cite{FootnoteSupplementary}. 
This large-scale, anisotropic forcing is represented in Fig.~\ref{Fig1}C as a small dark blue area elongated along $k_x$. 
After several seconds of shaking, in the inertial range our gas has a time-invariant, statistically isotropic momentum distribution, $\left\langle n({\bf k})\right\rangle\approx n(k) \propto k^{-\gamma}$, with $\gamma\approx 3.5$~\cite{navon2016emergence,FootnoteGamma}. This time invariance implies that the energy and particle fluxes through this $k$-range are $k$-independent, but it does not reveal their values. Here we extract the cascade fluxes by studying the dissipation in our gas.

In conventional fluids one observes macroscopic (hydrodynamic) degrees of freedom and the dissipation occurs in the form of heating, {\it i.e.} transfer of energy into the microscopic degrees of freedom. This dissipation is set by the viscosity $\nu$, which is generally not tuneable. Moreover, the resulting minute heating is often difficult to measure, due to thermal coupling of the fluid with its surroundings~\cite{cadot1997energy}.
Our system is thermally isolated from the environment and we have direct access to all the microscopic degrees of freedom, so the dissipation occurs only in the form of (readily measurable) particle loss. 
The optical box (Fig.~\ref{Fig1}B) has a non-infinite energy depth $U_\textrm{D}$, so particles with a sufficiently large energy leave the box; in momentum space $\UD$ corresponds to a sphere of radius $\kD=\sqrt{2m \UD}/\hbar$ (Fig.~\ref{Fig1}C), where $m$ is the atom mass.
This simple feature realises a synthetic dissipation scale, with $\UD$ defining the particle and energy sink. 
Crucially, this dissipation scale can be tuned by changing the trapping laser power \cite{FootnotekD}.

Formally, within the assumptions of the wave-turbulence theory, from the equations of motion one can derive a continuity equation with a source and a sink, that is local in momentum space~\cite{zakharov1992kolmogorov}:
\be\label{ContinuityEq}
\frac{\partial n({\bf k},t)}{\partial t}=F({\bf k},t)-D({\bf k},t) - \nabla_{\bf k}\cdot{\bf \Pi}_n ({\bf k},t) \, .
\ee
Here $F({\bf k},t)$ describes the forcing, $D({\bf k},t)$ the dissipation, and $\nabla_{\bf k} \cdot {\bf \Pi}_n$ the nonlinear interactions, where ${\bf \Pi}_n$ is the particle flux. For $F=D=0$, the steady-state solutions are zero-flux equilibrium thermodynamic states. 
If $F$ and $D$ are nonzero but are localised in $k$ space, one can also get non-equilibrium steady-state solutions with a nonzero scale-independent flux sustained by the source $F$ and the sink $D$.

For an isotropic outflow, the total radial particle flux is $\Pi_n (k) = 4\pi k^2 |{\bf \Pi}_n ({\bf k})|$.
Hence, from Eq.~(\ref{ContinuityEq}), in the inertial range $ 4\pi k^2 \, \partial n/\partial t = -\partial \Pi_n/\partial k$. Integrating over $k$ yields the intuitive result that we can measure the particle flux through the shell at $\kD$ by simply counting the atoms remaining in the trap (see Fig.~\ref{Fig1}C): 
\begin{equation}\label{EqParticleFlux}
\frac{\partial N}{\partial t} \equiv -\Pi_n(\kD,t)\, ,
\end{equation}
and for a (non-equilibrium) steady state, with time-invariant $n(k)$ in the inertial range~\cite{navon2016emergence}, the particle flux is $k$- and $t$-independent \cite{FootnoteFluxes1D3D}, so $\Pi_n(\kD,t) = \Pi_n(k,t)=\Pi_n$.

In steady state, the total radial energy flux, $\Pi_\mathcal{E}(k,t)$, is also $k$- and $t$-independent in the inertial range, and is equal to the rate of energy dissipation. To relate it to $\Pi_n$, we consider the pertinent case of weakly-interacting particles with a dispersion relation $\omega(k)$, so the energy spectrum is $\mathcal{E}(k,t)=\hbar\omega(k)n(k,t)$.
At $k < \kD$ microscopic interactions drive particles to both smaller and higher $k$, so the relationship between the {\it net} energy and particle fluxes, $\Pi_\mathcal{E}$ and $\Pi_n$, is nontrivial; one might naively expect that $\Pi_\mathcal{E}(k) = \hbar \omega(k) \, \Pi_n(k)$, but this cannot be true if both $\Pi_\mathcal{E}$ and $ \Pi_n$ are $k$-independent, while $\omega(k)$ is not. However, at $\kD$ the particles flow only one way, since there is no `back-flow' from the sink into the inertial range, so one can intuitively write
\begin{equation}\label{eq:twofluxes}
\Pi_\mathcal{E}(\kD) = \hbar \omega(\kD) \, \Pi_n(\kD) \, .
\end{equation}
Steady state then requires $\Pi_\mathcal{E} = \hbar \omega(\kD) \, \Pi_n$ at all $k$ in the inertial range; for our $\omega(k)$ this means that $\Pi_\mathcal{E} \propto \kD^2 \, \Pi_n$. Note that to formally derive Eq.~(\ref{eq:twofluxes}) one multiplies Eq.~(\ref{ContinuityEq}) by $\hbar\omega(k)$ and invokes the continuity equation for the energy to get
\begin{equation}\label{FluxesPartialLink}
\frac{\partial  \Pi_\mathcal{E} (k,t) }{\partial k} = \hbar \omega(k) \frac{\partial  \Pi_n (k,t)}{\partial k} \, 
\end{equation}
in the inertial range. For $k<\kD$ this equation is trivially satisfied by both of its sides being zero, and does not impose any relation between $\Pi_\mathcal{E}(k)$ and $\Pi_n(k)$. However, integrating it across a thin shell around $\kD$, and setting $n(k)$ and all fluxes to zero for $k>\kD$, recovers Eq.~(\ref{eq:twofluxes}).

\begin{figure}[t]
	\centerline{\includegraphics[width=\columnwidth]{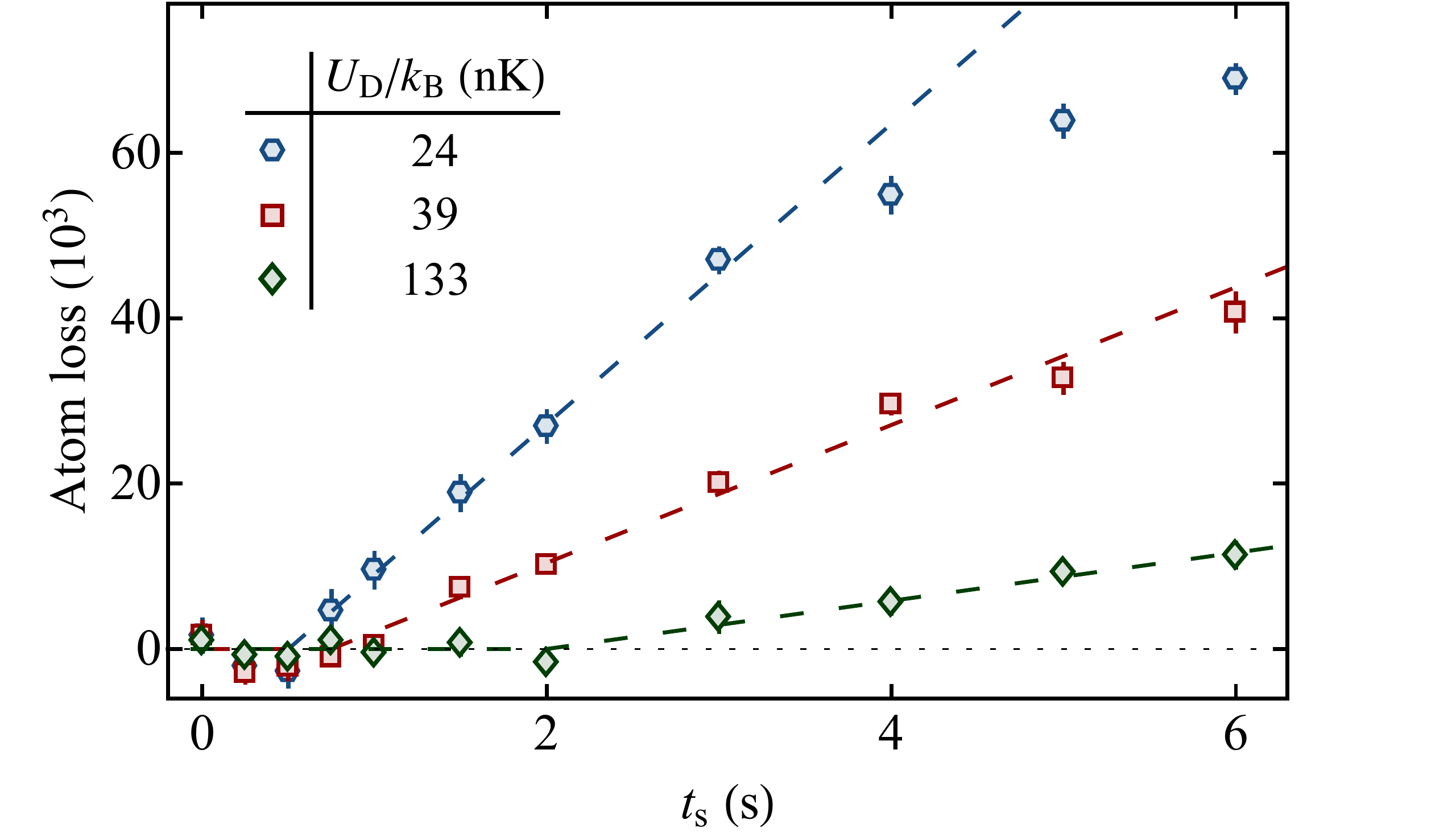}}
	\caption{{\bf Atom-loss dynamics due to the turbulent cascade.} Atoms lost versus shaking time $\ts$ for different trap depths $\UD$.
		Data points show averages of typically 50 measurements. Dashed lines are piece-wise linear fits. The systematic uncertainty in $\UD$ values is $20\%$.} 
	\label{Fig2}
\end{figure}

Experimentally, we vary $\kD$, while keeping $F_0$ fixed, and measure $\Pi_n(\kD)$ as per Eq.~(\ref{EqParticleFlux}). 
To mitigate the effects of the long-term few-percent drifts in the initial $N$, and of the additional atom loss through collisions with the background-gas particles, we perform differential measurements of the cascade-induced atom loss, $N_\textrm{loss}$, with reference measurements taken by setting  $F_0$ to zero in an otherwise identical experimental sequence.

In Fig.~\ref{Fig2}, we show $N_\textrm{loss}$ as a function of the shaking time $\ts$, for various values of $\UD$. 
In all cases at short times we observe no loss (within errors). This is consistent with the expectations that no losses occur at $k< \kD \propto \sqrt{\UD}$ and that initially it takes time for the excitations to cascade to $\kD$, when a steady state can be established (see Fig.~\ref{Fig1}). For $\ts$ longer than some onset time, $\tD$, the loss rate $\partial N_\textrm{loss}/\partial t$ is essentially constant in time, as long as the total loss is relatively small ($<30\%$ of the initial $N$). 
The dashed lines show piece-wise linear fits that we use to extract, for each $\UD$, both $\tD$ and the subsequent initial loss rate, which we identify with the steady-state particle flux $\Pi_n = \Pi_n(\kD)$. 
At much longer times, $\ts \gg \tD$, the steady-state assumption can no longer hold, because the losses significantly deplete the low-$k$ source of atoms.

\begin{figure}[t]
	\centerline{\includegraphics[width=\columnwidth]{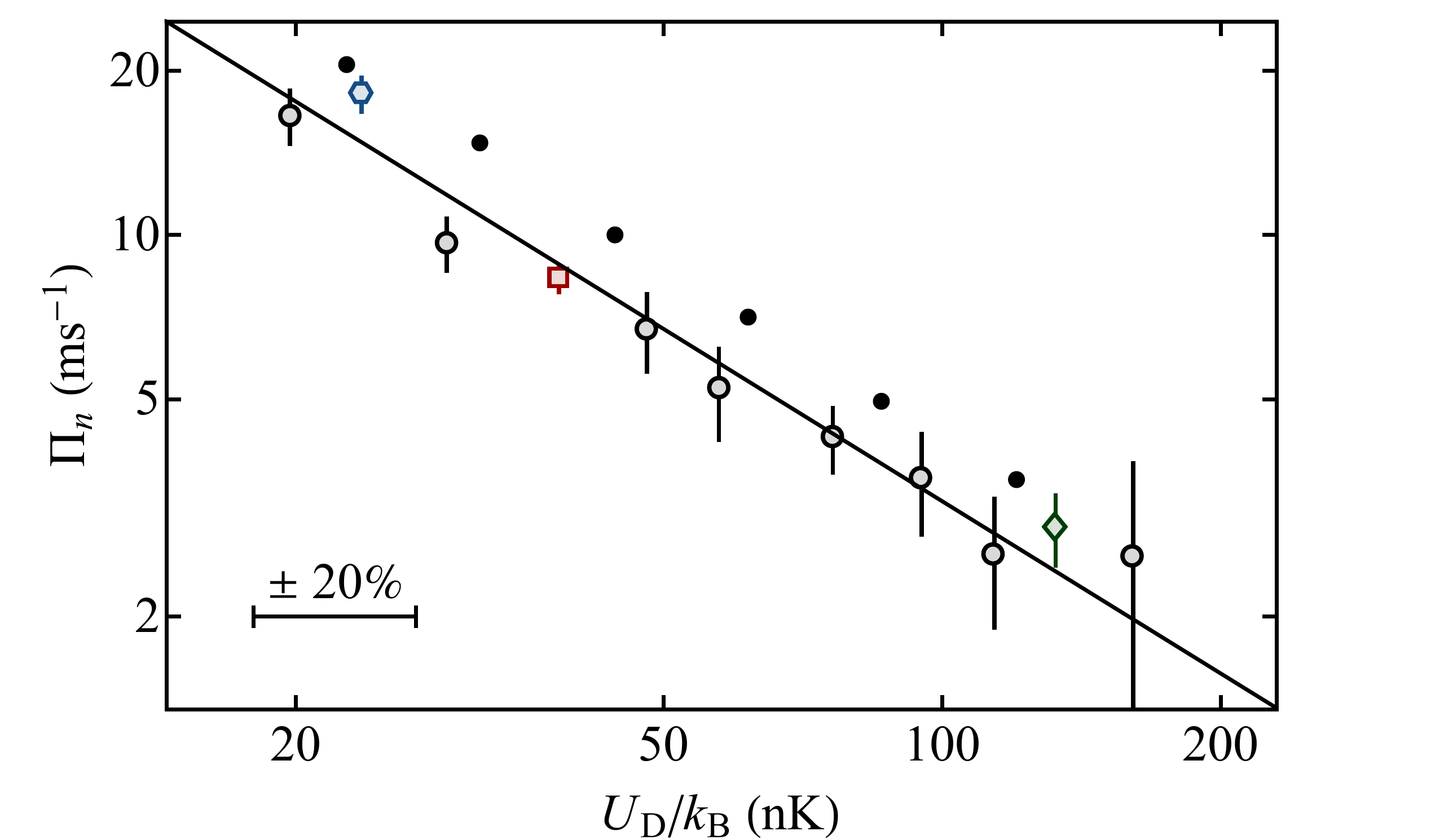}}
	\caption{{\bf Steady-state particle flux.} The atom-loss rate $\Pi_n$ versus the dissipation energy scale $\UD$ (open symbols), on a log-log plot; the three coloured points correspond to the data shown in Fig.~{\ref{Fig2}}. Solid symbols show the results of numerical simulations. The horizontal bar indicates the systematic uncertainty in the experimental $\UD$ values. A power-law fit to the experimental data (solid line) gives $\Pi_n \propto \UD^{-1.05(8)} \propto \kD^{-2.10(16)}$, in agreement with the theoretical prediction. 
	} \label{FigLossRate}
\end{figure}

In Fig.~\ref{FigLossRate}, we show a log-log plot of $\Pi_n$ versus $\UD$ \cite{FootnotekD}. We observe power-law behaviour $\Pi_n \propto \UD^{-1.05(8)} \propto \kD^{-2.10(16)}$. We complement these measurements with numerical simulations based on the Gross-Pitaveskii equation, for the same forcing protocol and without any free parameters (see~\cite{FootnoteSupplementary} for details). 
The numerical results are shown by solid circles; a fit to the numerical data (not shown) gives $\Pi_n \propto \UD^{-1.06(1)}$, in good agreement with the experimental data.

The so-called zeroth law of turbulence, first formulated in the context of classical incompressible fluids, stipulates that for fixed forcing the steady-state rate of energy dissipation tends to a nonzero constant as the viscosity vanishes ($\nu\rightarrow 0$) \cite{frisch1995turbulence,vassilicos2015dissipation}.
In our case, this corresponds to keeping $F_0$ fixed and taking $\kD \rightarrow \infty$ \cite{KolmogorovLengthScale}. This law implies that the particle flux should vanish as $\Pi_n \sim \kD^{-2}$ (see Eq.~(\ref{eq:twofluxes})), in excellent agreement with our data. Note that the steady-state energy balance also requires that $\Pi_\mathcal{E}$ is equal to the rate of energy input into the system, $\epsilon$. However, energy conservation alone is not sufficient to predict the scaling of $\Pi_n$ with $\kD$, because it is not {\it a priori} obvious that for fixed $F_0$ the rate at which the system absorbs energy from the drive is not affected by changing $\kD$~\cite{FootnoteZerothLaw}. 
{\it A posteriori}, we experimentally see that in our system the steady-state $\epsilon$ must be independent of the dissipation lengthscale down to our lowest $\kD$. 
Rather remarkably, if one changed $\kD$ dynamically, for a system to reach a new steady state the particle flux would have to self-consistently adjust at all $\kF < k < \kD$, since the steady-state $\Pi_n$ must be both $\kD$-dependent (to satisfy the zeroth law) and $k$-independent for a given $\kD$.

\begin{figure}[t]
	\centerline{\includegraphics[width=\columnwidth]{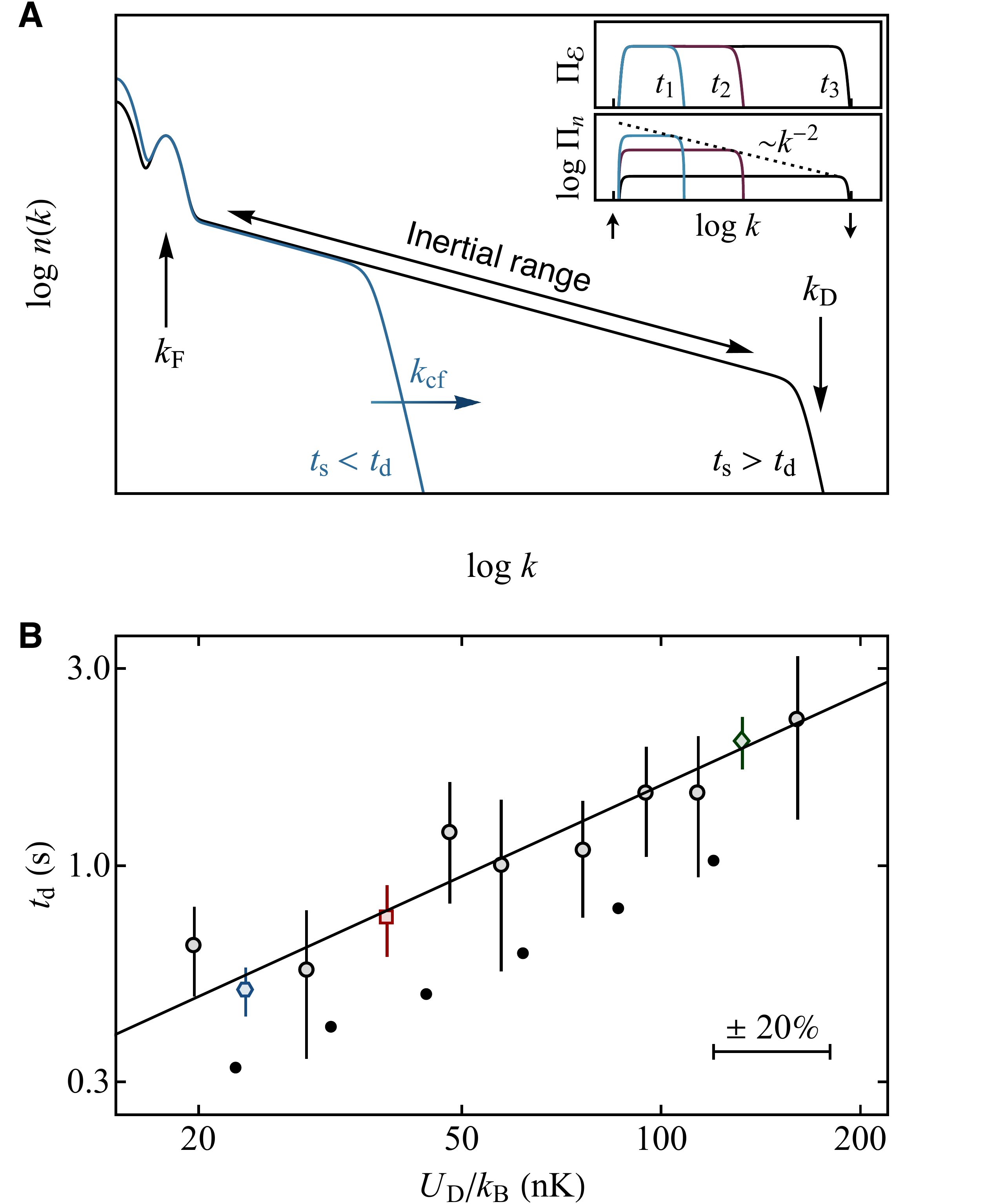}}
	\caption{{\bf Establishing the steady state: the cascade-front dynamics.} 
		({\bf A}) Momentum-space turbulent dynamics. Forcing occurs at $\kF$ and the steady-state distribution $n(k)$ is established in the wake of the cascade front $\kf(\ts)$, which propagates outwards until it reaches $\kD$ at time $\tD$. Inset: Consistent picture for the evolution of the energy flux $\Pi_\mathcal{E}$ and particle flux $\Pi_n$, for three different times $t_1$ (blue) $<t_2$ (purple) $<t_3$ (black), with $t_2 < \tD < t_3$. The forcing and dissipation scales are indicated by the vertical arrows, as in the main panel. ({\bf B}) Onset time for dissipation. Open symbols show the measured  $\tD$ values versus $\UD$ on a log-log plot; the three coloured points correspond to the data shown in Fig.~{\ref{Fig2}}. Solid symbols show the results of numerical simulations. 
		The horizontal bar indicates the systematic uncertainty in the experimental $\UD$ values.
		A power-law fit, $\tD \propto \UD^{\beta}$, to the experimental data (solid line) gives $\beta = 0.73(6)$, in agreement with the prediction $\beta = 0.75(5)$.
	} \label{FigFront}
\end{figure}

Having established a consistent picture of the steady-state fluxes in our gas, we now turn to the pre-steady-state turbulent dynamics. 
In Fig.~\ref{FigFront}A, we outline a consistent picture of the early-time dynamics in Fourier space.  
The forcing, which generates a surplus of atoms at $\kF$, initiates the cascade at $\ts =0$. As the cascade front, $\kf(\ts)$, propagates to higher $k$, the steady-state momentum distribution, $n(k) \propto k^{-\gamma}$, is established in its wake (see also~\cite{FootnoteSupplementary}). The dynamics is dissipationless until $\kf$ reaches $\kD$, at time $\tD$; only then a steady state, with matching $\epsilon$ and $\Pi_\mathcal{E}(\kD)$, is established. Hence, the fact that we can experimentally observe the initial dissipationless stage of turbulence ($\ts <\tD$), and the dependence of $\tD$ on $\UD$, gives us access to the dynamics of the cascade front in momentum space.

At $\ts < \tD$, the instantaneous particle flux is $k$-independent for $k<\kf (\ts)$, vanishes for $k>\kf (\ts)$, and must match the rate of the population increase in the inertial range: $n(\kf) \, 4\pi \kf^2 \, \textrm{d}\kf = \Pi_n(\kf)  \, \textrm{d}\ts$, so $\kf^{2-\gamma} \, \textrm{d}\kf \propto \Pi_n(\kf)  \, \textrm{d}\ts$. Analogously, for the increase of total energy in the inertial range, $\kf^{4-\gamma} \textrm{d}\kf \propto \Pi_\mathcal{E}(\kf)  \, \textrm{d}\ts$, and $\Pi_\mathcal{E}(\kf)$ is equal to the instantaneous energy-injection rate $\epsilon$.

Assuming that $\epsilon$, which we found not to depend on $\kD$ in steady state, is also independent of $\kf$ at $\ts < \tD$, we get that the instantaneous $\Pi_n (\ts)$, at $k<\kf(\ts)$, is $\propto \kf^{-2}$. This gives an elegant unifying picture of the particle fluxes for $\ts < \tD$ and $\ts > \tD$ (see the inset of Fig.~\ref{FigFront}A): $\Pi_n$ is always the same function of the highest $k$  for which the steady-state $n(k)$ has been established ({\it i.e.} the lowest $k$ from which there is no back-flow), whether that is the instantaneous $\kf < \kD$ (for $\ts < \tD$) or $\kD$. This self-consistent picture also leads to a quantitative prediction that is verifiable in our experiments: 
the time independence of $\epsilon$ implies $\kf^{4-\gamma} \textrm{d}\kf \propto \textrm{d}\ts$, which for $\gamma < 5$ and $\kD \gg \kF$ gives a power-law prediction $\tD \propto \UD^{\beta}$, with  $\beta = (5-\gamma)/2$. 
Specifically, for our $\gamma=3.5(1)$~\cite{navon2016emergence}, we predict $\beta = 0.75(5)$.

In Fig.~\ref{FigFront}B, we show the variation of $\tD$ with $\UD$. We find that our data is indeed well described by a power-law, with $\beta = 0.73(6)$, in agreement with our prediction.
We again also show the results of our numerical simulations (solid circles), which show similar behaviour with a small systematic offset; a fit to the numerical data (not shown) gives $\beta = 0.68(2)$.

Finally, it is interesting to note that the criterion for $\tD$ to show scaling behaviour, namely $\gamma < 5$ and hence $\beta>0$, is intimately linked to another important concept in the theory of turbulence. For $\gamma <5$ the steady-state spectrum has infinite capacity, meaning that it carries infinite energy for $\kD \rightarrow \infty$.  
It is indeed generally expected for infinite-capacity systems that the cascade front propagates at a finite speed and that the Kolmogorov-Zakharov turbulence spectrum forms right behind it~\cite{nazarenko2011wave}. 
It is also important to note that for $\beta >0$, in the limit $\kD \rightarrow \infty$ the steady state is actually never reached, since $\tD \rightarrow \infty$. This reinforces the fact that this limit is not experimentally meaningful, and one can recover formal theoretical statements only by
varying $\kD$.

Our work establishes a qualitatively new view on wave turbulence, providing a complete consistent picture of the dynamics at both short (pre-steady-state) and long (steady-state) times. The possibility of synthetic dissipation also opens new theoretical perspectives. In the future it would be interesting to engineer arbitrary momentum-cutoff landscapes, which could, for example, allow studies of anisotropic turbulence. By dynamically tuning the dissipation scale, or the driving force, it should also be possible to study quenches between different turbulent states.  

We thank Ehud Altman, Dan Stamper-Kurn, Frédéric Chevy, and Jake Glidden for discussions, and Timon Hilker for comments on the manuscript. This work was supported by EPSRC [Grants No. EP/N011759/1 and No. EP/P009565/1], ERC (QBox), QuantERA (NAQUAS, EPSRC Grant No. EP/R043396/1), AFOSR, and ARO. N.N acknowledges support from the David and Lucile Packard Foundation. A.L.G. and N.N. acknowledge support from Trinity College (Cambridge). R.L. acknowledges support from the E.U. Marie-Curie program [Grant No. MSCA-IF-2015 704832] and Churchill College, Cambridge.  R.P.S. acknowledges support from the Royal Society. K.F. was supported by JSPS KAKENHI Grant No. JP16J01683. M. T. acknowledges support from JSPS KAKENHI Grant No. 17K05548 and MEXT KAKENHI Grant No. 16H00807.


\section*{SUPPLEMENTARY MATERIAL}

\setcounter{section}{0}
\setcounter{subsection}{0}
\setcounter{figure}{0}
\setcounter{equation}{0}

\renewcommand{\figurename}[1]{Fig.}
\makeatletter
\renewcommand{\thefigure}{S\@arabic\c@figure} 
\makeatother
\renewcommand\thetable{\arabic{table}}
\renewcommand{\vec}[1]{{\boldsymbol{#1}}}

\renewcommand{\theequation}{{S}\arabic{equation}}
\newcommand{\equref}[1]{equation \ref{#1}}
\makeatletter
\DeclareRobustCommand{\gobblesomeargs}[2]{#2}
\renewcommand{\p@equation}{S\gobblesomeargs}
\makeatother

\section{Calibration of the shaking force and the resonant driving frequency}

The shaking force is produced by coils that create a magnetic field gradient. We calibrate its magnitude $F$ (for a given voltage applied to the coils) by switching off the box trap, immediately pulsing the force for a time $\delta t$, and measuring the resulting velocity kick $\delta v = F \delta t/m$; to determine $\delta v$ we measure the position of the cloud's centre of mass, $x_\textrm{CoM}$, after a time of flight $t_\textrm{ToF}$. 

\begin{figure}[h]
	\centerline{\includegraphics[width=\columnwidth]{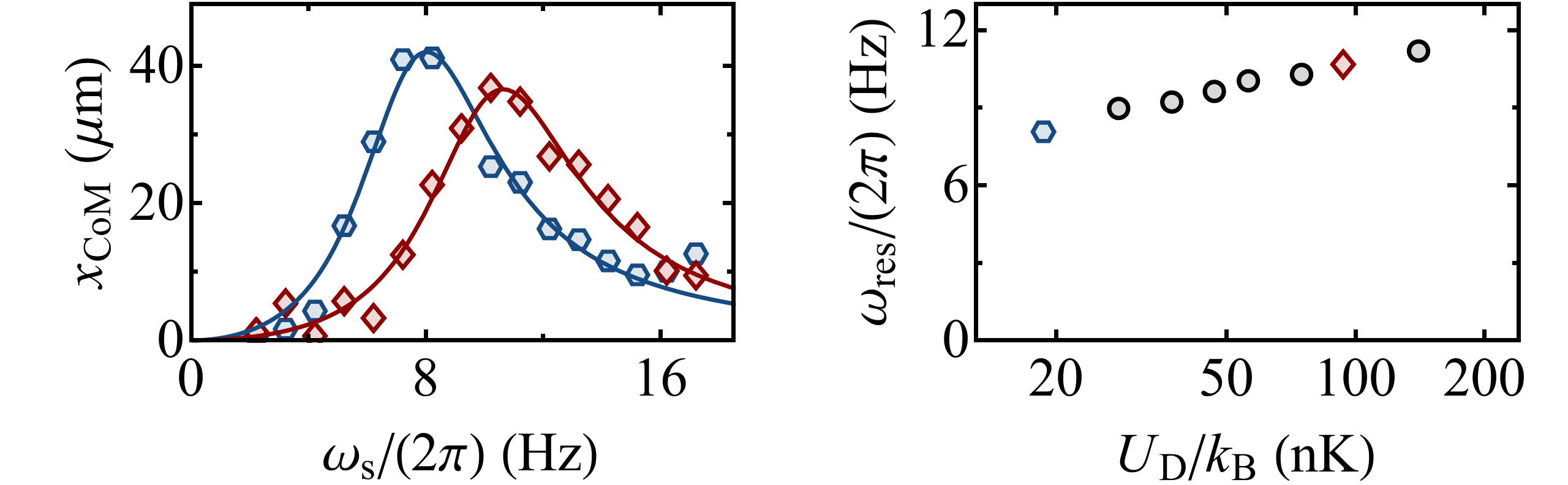}}
	\caption{{\bf The resonant drive frequency.} Left: Resonance measurements for $\UD/k_{\rm B}=19$~nK (blue) and $94$~nK (red); solid lines are fits based on Eq.~(\ref{eq:omega}). Right: $\omega_\textrm{res}$ versus $\UD$.
		\label{FigFreqCalib}}
\end{figure}

Due to the optical resolution of the system used to create the box trap, the trap walls are not perfectly sharp~\cite{gaunt2013boseSI}, and consequently the frequency of the lowest axial sound mode, $\omega_\textrm{res}$, slightly depends on $\UD$. To ensure that the gas is always driven on resonance, we measure $\omega_\textrm{res} (\UD)$. We perform stroboscopic modulation spectroscopy by applying the driving force $F_0\sin(\omega_\textrm{s} t)$ for $t_\textrm{s}=2$ s, with $F_0 L\approx \kB \times 2.5$~nK, and then releasing the cloud and measuring $x_\textrm{CoM}$ after $t_\textrm{ToF}=140$ ms. Choosing discrete values of $\omega_\textrm{s}$ such that $\omega_\textrm{s} t_\textrm{s}=2 \pi j+\pi/2$, where $j$ is an integer, the resulting $x_\textrm{CoM}$ has an absorptive shape:
\begin{equation}
x_\textrm{CoM} \propto \frac{\Gamma^2 \omega_\textrm{s}^2}{(\omega_\textrm{s}^2-\omega_\textrm{res}^2)^2+\Gamma^2 \omega_\textrm{s}^2} \, ,
\label{eq:omega}
\end{equation}
where $\Gamma$ is the linewidth. In Fig.~\ref{FigFreqCalib} we show such line shapes for two different $\UD$, and the plot of $\omega_\textrm{res}$ versus $\UD$.  

\section{Numerical simulations}

\subsection{Gross-Pitaevskii simulations with dissipation}

The starting point for our simulations is the Gross-Pitaevskii equation (GPE) for the classical field $\psi({\bf r},t)$:
\begin{eqnarray} 
i\hbar \frac{\partial\psi}{\partial t} = \left(-\frac{\hbar^2}{2m} \nabla^2 + V({\bf r},t) + g|\psi|^2 \right)\psi \, ,
\label{GP}
\end{eqnarray}
where $g=4\pi\hbar^2 a/m$ and $a$ is the $s$-wave scattering length. If $V({\bf r},t)\in \mathbb{R}$, the evolution of the GPE conserves the total particle number $N=\int |\psi|^2 \, \textrm{d}{\bf r}$. 
Introducing dissipation in the GPE is a subtle problem~\cite{kobayashi2005kolmogorov,griffin2009bose,proment2009quantum,reeves2015identifying}. We introduce a phenomenological term in $V({\bf r},t)$ that closely resembles the dissipation process in our experiment. We write: 
\begin{eqnarray} \nonumber
V({\bf r},t) = V_{\rm box}({\bf r}) + V_{\rm osc}({\bf r},t) - i V_{\rm diss}({\bf r}),  \label{Pot}
\end{eqnarray}
where $V_{\rm box}$ is the box potential, $V_{\rm osc}$ is the forcing potential, and $i V_{\rm diss}$ is an imaginary `sponge' potential that `absorbs' particles with sufficiently high energy to leave the trap and removes them from the system. More precisely: 
\begin{eqnarray} \nonumber
V_{\rm box}({\bf r}) = \begin{cases}
0& \rm{if } \, |{\it x}| \leq \frac{\it{L}}{2}, \sqrt{{\it y}^2+{\it z}^2} \leq \it{R}\\
\UD  & {\rm otherwise} \, , 
\end{cases}
\end{eqnarray}
\begin{eqnarray} \nonumber
V_{\rm osc}({\bf r},t) =  F x~{\rm sin }(\omega_{\rm res} t) \, , 
\label{exc}
\end{eqnarray}
and
\begin{eqnarray} \nonumber
V_{\rm diss}({\bf r})= \begin{cases}
0& \rm{if } \, |{\it x}| \leq \frac{{\it L}+2\delta}{2}, \sqrt{{\it y}^2+{\it z}^2} \leq \it{R}+\delta  \\
V_{\rm D}  & {\rm otherwise} \, .
\end{cases}
\end{eqnarray}
The phenomenological parameter $\delta$, the spatial offset between the edge of the box and the sponge, is introduced because even if all particles are trapped, for a non-infinite $\UD$ an evanescent component of $\psi({\bf r},t)$ exists outside the box. We have verified that for a wide range of $V_{\rm D}$ and $\delta$ our results do not depend on their exact values (see Fig.~\ref{FigSimulations} below).

We numerically solve Eq.~(\ref{GP}) using a pseudo-spectral method with the fourth-order Runge-Kutta time evolution.
In simulations $L=27~\mu$m, $R=16~\mu$m, and the initial atom number is $N_0 = 1.1\times 10^5$, corresponding to chemical potential $\mu=gn_0 =\kB \times 1.9~{\rm nK}$, where $n_{0}=N_0/(\pi R^2L)$. 
The size of our whole numerical grid is $40 \, \xi \times 40 \, \xi \times 40 \, \xi$, where $\xi=\hbar/\sqrt{2mgn_0} = 1.2~\mu$m is the healing length. The spatial and temporal resolutions are $\frac{40}{128}\xi$ and $10^{-3}\hbar/\mu = 4.1~\mu{\rm s}$ respectively. 

The initial $\psi({\bf r},t=0)$ is determined by calculating the ground state in the static trap ($F=0$ and $V_{\rm D}=0$), using imaginary-time evolution of the GPE. The resonant driving frequency $\omega_{\rm res}$ is then determined by numerically solving the Bogoliubov equations on $\psi({\bf r},0)$. For all experimentally explored $\UD$ we get $\omega _{\rm res} \approx 2\pi\times 8.5$~Hz, with variations of $< 3\%$. 
Finally, to simulate the shaking experiments, we solve the real-time GPE with the forcing amplitude $F = F_0=1.36~\mu/L$ and nonzero $V_{\rm D}$.

\subsection{Atom-loss dynamics}

\begin{figure}[h]
	\centerline{\includegraphics[width=1\columnwidth]{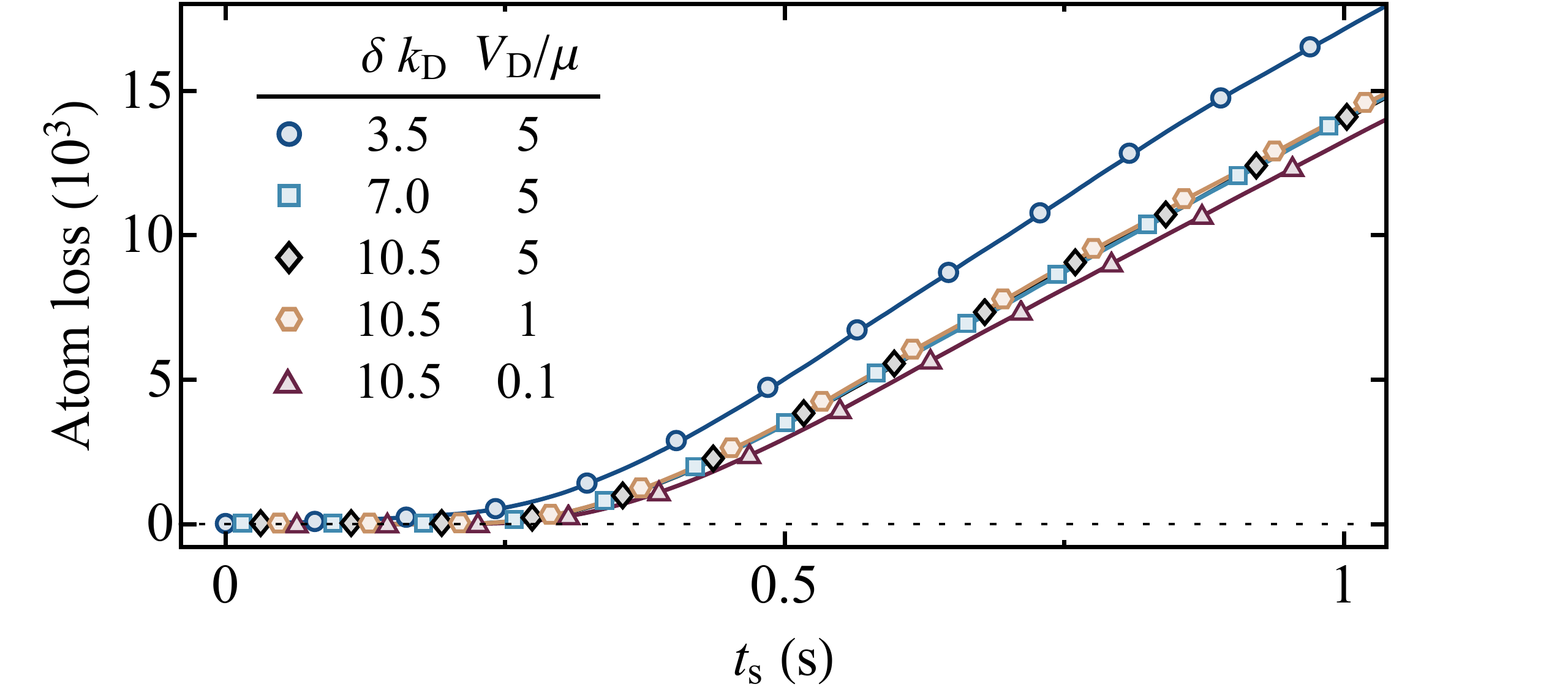}}
	\caption{{\bf Atom-loss dynamics in numerical simulations.}  Atoms lost versus shaking time for $\UD=\kB\times 23$~nK and various combinations of the dissipation parameters $V_{\rm D}$ and $\delta$.}
	\label{FigSimulations}
\end{figure}

In Fig.~\ref{FigSimulations} we show simulated atom loss $N_0 - N(t_\textrm{s})$, where $N(t_\textrm{s}) = \int |\psi({\bf r},t_\textrm{s})|^2  \, \textrm{d}{\bf r}$, for $\UD=\kB\times 23$~nK and various  combinations of the dissipation parameters $V_{\rm D}$ and $\delta$. In all cases we see curves similar to the experimental ones shown in Fig. 2 in the main paper (and for the results shown in the main paper we analyse them in the same way as the experimental data). For a fixed $V_{\rm D} =5 \mu$, we get essentially indistinguishable results for any $\delta \gtrsim 7/\kD$. Qualitatively, $\delta$ needs to be sufficiently larger than $1/\kD$ for the probability of absorbing (on a timescale $t_\textrm{s}$) particles with energies below $\UD$ to be vanishingly small; otherwise we remove too many particles. For a fixed $\delta = 10.5/\kD$ we get essentially the same results for any $V_{\rm D} \gtrsim \mu$. 

\subsection{Fourier-space dynamics}

\begin{figure}[h!]
	\centerline{\includegraphics[width=1\columnwidth]{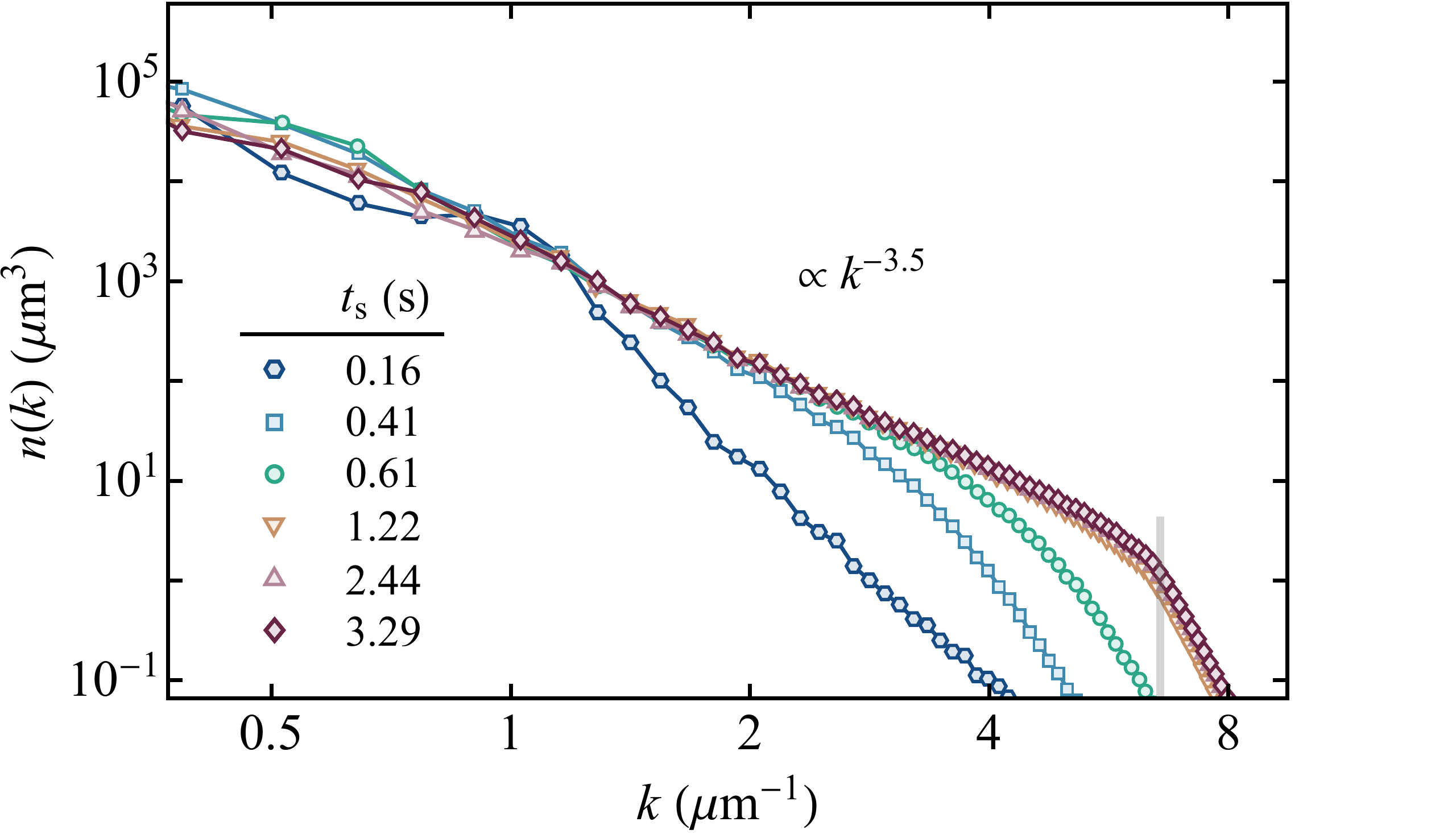}}
	\caption{{\bf Momentum-space dynamics in numerical simulations.}  $n(k)$ computed for $\UD=\kB\times 120$~nK and various shaking times $t_\textrm{s}$. The power-law momentum distribution $n(k)\sim k^{-\gamma}$, with $\gamma \approx 3.5$, develops in the wake of the cascade front. A steady state is established once the cascade front reaches $\kD$; in this example  $\kD=8/\xi$, indicated by the vertical grey band.  
		\label{Fignk}}
\end{figure} 

We also compute the evolution of the momentum distribution in the presence of shaking and dissipation, supporting the qualitative picture outlined in Fig.~4A in the main paper. 
The momentum distributions are averaged over spherical shells to obtain $n(k)$, and normalised such that $\sum_k 4\pi k^2  n(k)\delta k=N$, where $\delta k=\frac{\pi}{20\xi}$ is the grid resolution in ${\bf k}$ space.
In Fig.~\ref{Fignk}, we show $n(k)$ for $\UD=\kB\times 120$~nK and various shaking times $t_\textrm{s}$. The power-law distribution $n(k)\sim k^{-3.5}$ develops in the wake of the cascade front, and a steady state is established in the inertial range once the cascade front reaches the dissipation scale (grey band).


\begin{thebibliography}{10}
\bibitem{FootnotekD}
	Note that the lowest $\UD$ we explore corresponds to $\UD/\mu\approx 10$,
	$\UD/(F_0 L)\approx 8$ and $\kD/\kF\approx 23$.
\bibitem{KolmogorovLengthScale}
	For incompressible flows, the Kolmogorov dissipation lengthscale, analogous to our $1/\kD$, depends on the viscosity of the fluid (as $\propto \nu^{3/4}$).
\bibitem{FootnoteZerothLaw}
	In most familiar systems, like the damped harmonic oscillator, the rate at which the system absorbs energy in steady state depends on both the driving and the dissipation.
\end{thebibliography}
\end{document}